\documentclass[twocolumn,a4paper,amsmath,amssymb,showpacs,prb,superscriptaddress]{revtex4-1}
 
\usepackage{graphicx}% Include figure files
\usepackage{amssymb, amsfonts, amsmath}
\usepackage{dcolumn}% Align table columns on decimal point
\usepackage{bm}% bold math
\usepackage{color}
\usepackage[normalem]{ulem}

\begin{document}

\title{Cluster multipole theory for anomalous Hall effect in antiferromagnets}

\author{M.-T. Suzuki}
\affiliation{RIKEN Center for Emergent Matter Science (CEMS), Wako, Saitama
351-0198, Japan}
\author{T. Koretsune}
\affiliation{RIKEN Center for Emergent Matter Science (CEMS), Wako, Saitama
351-0198, Japan}
\affiliation{JST, PRESTO, 4-1-8 Honcho Kawaguchi, Saitama 332-0012, Japan.}
\author{M. Ochi}
\affiliation{Department of Physics, Osaka University, Toyonaka, Osaka 560-0043, Japan}
\author{R. Arita}
\affiliation{RIKEN Center for Emergent Matter Science (CEMS), Wako, Saitama
351-0198, Japan}
\date{\today}

\begin{abstract}
  We introduce a cluster extension of multipole moments to discuss the
 anomalous Hall effect (AHE) in both ferromagnetic (FM) and
 antiferromagnetic (AFM) states in a unified framework.
  We first derive general symmetry requirements for the AHE in the
 presence or absence of the spin-orbit coupling,
 by considering the symmetry of the Berry curvature in ${\bm k}$ space.
 The cluster multipole (CMP) moments are then defined to quantify the
 macroscopic magnetization in non-collinear AFM states, 
 as a natural generalization of the magnetization in FM states.
  We identify the macroscopic CMP order which induces the AHE.
  The theoretical framework is applied to the non-collinear
 AFM states of Mn$_3$Ir, for which an AHE was predicted in a
 first-principles calculation, and Mn$_3$$Z$ ($Z$=Sn, Ge), for which a
 large AHE was recently discovered experimentally.
  We further compare the AHE in Mn$_3$$Z$ and bcc Fe in terms of the CMP.
  We show that the AHE in Mn$_3$$Z$ is characterized with
   the magnetization of a cluster octupole moment in the same
 manner as that in bcc Fe characterized with the magnetization of the dipole moment.
\end{abstract}

%\pacs{}
%72.15.Qm : Scattering mechanisms and Kondo effect
%74.62.Dh : Effects of crystal defects, doping and substitution
%74.70.Tx : Heavy-fermion superconductors
%81.15.Hi : Molecular, atomic, ion, and chemical beam epitaxy

\maketitle

\section{INTRODUCTION}
 The modern formalism of the intrinsic anomalous Hall conductivity (AHC)
 provides profound insight into the anomalous Hall effect (AHE) of being
 closely related with the topology of one-electron energy
 bands~\cite{Nagaosa2010,Xiao2010, Gradhand2012}.
  The AHE is usually observed in ferromagnetic (FM) metals, but the
 AHE has been studied also for certain non-collinear
 antiferromagnetic (AFM) states~\cite{Yoshii2000, Taguchi2001, Shindou2001, Yoshii2006, Tomizawa2009,
 Nagaosa2010, Tomizawa2010, Taguchi2003, Taguchi2004, Chen2014, Kuebler2014}.
  Especially, a large AHC has been recently discovered for the AFM
 states in Mn$_3$$Z$ ($Z$=Sn, Ge), whose magnetic geometry has no
 uniform magnetization~\cite{Nakatsuji2015, Kiyohara2016, Nayak2016}.
 The topological feature of these AFM states has also been
 investigated based on the first-principles calculations~\cite{Chen2014,
 Kuebler2014, Yang2016}.

  The AHE requires not only the broken time-reversal symmetry
 but also a certain type of magnetic
 structure~\cite{Kleiner1966,Seemann2015}. Furthermore, the situation
 changes depending on the presence or absence of the spin-orbit (S-O) 
 coupling. For instance, the AHE requires the S-O coupling in collinear
 FM states, characterized by uniform magnetization.
  The crucial role of the S-O coupling for the AHE in the FM states has been
 discussed since its pioneering work\cite{Karplus1954}. 
   On the other hand, in AFM systems, two types of AHE have been investigated. 
  One is the AHE in non-coplanar spin configurations where the AHE can be
 induced even without the S-O coupling.
  The AHE is characterized by the scalar spin chirality~\cite{Note5} and
 studied intensively in the context of the topological Hall
 effect~\cite{Shindou2001,Nagaosa2013}.
 The other is the AHE in coplanar spin systems such as
 Mn$_3$Sn~\cite{Chen2014,Kuebler2014,Zhang2016}. In this case,
 it is not well understood whether there is a macroscopic quantity that
 characterizes the AHE such as the uniform magnetization or scalar spin
 chirality.
Moreover, there is no clear explanation for what types of AFM structures
 induce the AHE.

   The purpose of this paper is to provide comprehensive understanding of
 the AHE in relation to the magnetic structure.
  We propose a new order parameter, which we call the cluster
 multipole (CMP) moment, to measure the symmetry breaking of
 commensurate non-collinear magnetic order.
 This systematically explains what types of AFM structures induce
  the AHE and whether the AHE requires the
 S-O coupling or not in that AFM state.

 The structure of this paper is as follows.
   In Sec. \ref{Sec:Sym}, we derive a symmetry condition for finite
   AHC in generic non-collinear magnetic systems by considering the symmetry of the Berry
 curvature in ${\bm k}$ space. We show that the AHE
   is forbidden to emerge by some symmetry elements of the magnetic space group,
   whose operations preserve the magnetic structure.
 The derivation also leads to comprehensive understanding for the
 requirement of the S-O coupling for the AHE. 
  In Sec. \ref{Sec:Multipole}, we introduce CMP moments
   as order parameters defined for a cluster of atoms,
   which is a natural generalization of the local magnetic moments for atoms.
   The CMP characterizes the non-collinear AFM structure as
   analogous to the atomic magnetic multipole moments characterizing
   the local magnetic distribution~\cite{Kusunose2008, Kuramoto2008,
   Kuramoto2009, Santini2009, Suzuki2010, Suzuki2013}.
  We show that the AHE of Mn$_3$Ir and Mn$_3$$Z$ is associated with the cluster
   octupole moments which belong
 to the same symmetry with the magnetic dipole moments.
  In Sec. \ref{Sec:FirstPrinciples}, we calculate the electronic structure, Berry
   curvature and AHC for the AFM states of Mn$_3$$Z$ from first principles.
 As a reference, we also calculated
 those properties for the FM state of bcc-Fe, which has well been
 investigated in earlier works~\cite{Wang2006, Gosalbez2015, Fang2003}.
   We show that the AHE of the FM and AFM states can be discussed in
   the same framework in terms of the CMP.
 Finally, a summary of the results is given in Sec. \ref{Sec:Summary}.
\\
\\
%%%%%%%%%%%%% Symmetry of Berry curvature %%%%%%%%%%%%%%%
\begin{widetext}
  \begin{center}
  \begin{table}[tb]
    \caption{Constraint on the Berry curvature in ${\bm k}$ space for
   some representative symmetries. The $x$, $y$, $z$ express the
   Cartesian coordinates. $C_{n\mu}$ indicates the $n$-fold rotation
   operator along the $\mu$-axis, $P$ is the spacial inversion operator,
   and $T$ is the time reversal operator.
   Mirror operation whose mirror plane normal to the $\mu$-axis
   corresponds to $PC_{2\mu}$.}
   \begin{tabular}{cccc} \hline \hline
 \multicolumn{2}{c}{Unitary operators}  &  \multicolumn{2}{c}{Anti-unitary operators} \\
\hline
        & $\Omega^{x}(-k_{x},-k_{y},k_{z})= -\Omega^{x}(k_{x},k_{y},k_{z})$  &          
    & $\Omega^{x}(k_{x},k_{y},-k_{z}) = \Omega^{x}(k_{x},k_{y},k_{z})$ \\
 $C_{2z}$ &    $ \Omega^{y}(-k_{x},-k_{y},k_{z}) =
	-\Omega^{y}(k_{x},k_{y},k_{z})$  &  $T C_{2z}$ &
    $\Omega^{y}(k_{x},k_{y},-k_{z}) = \Omega^{y}(k_{x},k_{y},k_{z})$ \\
          &  $ \Omega^{z}(-k_{x},-k_{y}, k_{z}) =   \Omega^{z}(k_{x},k_{y}, k_{z})$ &
	    &  $\Omega^{z}(k_{x},k_{y},-k_{z}) = -\Omega^{z}(k_{x},k_{y},k_{z})$ \\
\hline
           &  $\Omega^{x}( k_{x},k_{y},-k_{z})=-\Omega^{x} (k_{x},k_{y},k_{z})$ &
           &  $\Omega^{x}( -k_{x}, -k_{y},k_{z})=\Omega^{x} (k_{x},k_{y},k_{z})$\\
  $PC_{2z}$ &  $\Omega^{y}( k_{x},k_{y},-k_{z})=-\Omega^{y}(k_{x},k_{y},k_{z})$ 
           & $T PC_{2z}$ &  $\Omega^{y}( -k_{x}, -k_{y},k_{z})=\Omega^{y}( k_{x}, k_{y},k_{z})$\\
           & $\Omega^{z}(k_{x},k_{y},-k_{z}) = \Omega^{z}( k_{x}, k_{y},k_{z})$ &
           & $\Omega^{z}( -k_{x}, -k_{y},k_{z})=-\Omega^{z}( k_{x}, k_{y},k_{z})$\\
\hline
    & $\Omega^{x}(-k_{y},-k_{x},-k_{z})= -\Omega^{y}(k_{x},k_{y},k_{z})$  &          
    & $\Omega^{x}(k_{y},k_{x},k_{z}) = \Omega^{y}(k_{x},k_{y},k_{z})$ \\
 $C_{2[1{\bar 1}0]}$ &  $\Omega^{y}(-k_{y},-k_{x},-k_{z}) =
	-\Omega^{x}(k_{x},k_{y},k_{z})$  
&  $T C_{2[1{\bar 1}0]}$ & $\Omega^{y}(k_{y},k_{x},k_{z}) = \Omega^{x}(k_{x},k_{y},k_{z})$ \\
            &  $\Omega^{z}(-k_{y},-k_{x},-k_{z}) =-\Omega^{z}(k_{x},k_{y}, k_{z})$ &
	    &  $\Omega^{z}(k_{y},k_{x},-k_{z}) =  \Omega^{z}(k_{x},k_{y},k_{z})$ \\
\hline
    & $\Omega^{x}(k_{y},k_{x},k_{z})= -\Omega^{y}(k_{x},k_{y},k_{z})$  &          
    & $\Omega^{x}(-k_{y},-k_{x},-k_{z}) = \Omega^{y}(k_{x},k_{y},k_{z})$ \\
 $PC_{2[1{\bar 1}0]}$ &  $\Omega^{y}(k_{y},k_{x},k_{z}) =
	-\Omega^{x}(k_{x},k_{y},k_{z})$  
&  $T PC_{2[1{\bar 1}0]}$ & $\Omega^{y}(-k_{y},-k_{x},-k_{z}) = \Omega^{x}(k_{x},k_{y},k_{z})$ \\
            &  $\Omega^{z}(k_{y},k_{x},k_{z}) =-\Omega^{z}(k_{x},k_{y}, k_{z})$ &
	    &  $\Omega^{z}(-k_{y},-k_{x},-k_{z}) =  \Omega^{z}(k_{x},k_{y},k_{z})$ \\
\hline
             & $\Omega^{x}(k_{z},k_{x},k_{y})=\Omega^{y}(k_{x},k_{y},k_{z})$ &  &$\Omega^{x}(-k_{z},-k_{x},-k_{y})=-\Omega^{y}(k_{x},k_{y},k_{z})$  \\
$C_{3[111]}$ &$\Omega^{y}(k_{z},k_{x},k_{y})=\Omega^{z}(k_{x},k_{y},k_{z})$
        & $T C_{3[111]}$ & $\Omega^{y}(-k_{z},-k_{x},-k_{y})=-\Omega^{z}(k_{x},k_{y},k_{z})$ \\
        & $\Omega^{z}(k_{z},k_{x},k_{y})=\Omega^{x}(k_{x},k_{y},k_{z})$ & &$\Omega^{z}(-k_{z},-k_{x},-k_{y})=-\Omega^{x}(k_{x},k_{y},k_{z})$  \\

\hline
             & $\Omega^{x}(-k_{z},-k_{x},-k_{y})=\Omega^{y}(k_{x},k_{y},k_{z})$ &  &$\Omega^{x}(k_{z},k_{x},k_{y})=-\Omega^{y}(k_{x},k_{y},k_{z})$  \\
$PC_{3[111]}$ &$\Omega^{y}(-k_{z},-k_{x},-k_{y})=\Omega^{z}(k_{x},k_{y},k_{z})$
        & $T PC_{3[111]}$ & $\Omega^{y}(k_{z},k_{x},k_{y})=-\Omega^{z}(k_{x},k_{y},k_{z})$ \\
        & $\Omega^{z}(-k_{z},-k_{x},-k_{y})=\Omega^{x}(k_{x},k_{y},k_{z})$ & &$\Omega^{z}(k_{z},k_{x},k_{y})=-\Omega^{x}(k_{x},k_{y},k_{z})$  \\
\hline \hline
   \end{tabular}
\label{tab:BerryTransform}
  \end{table}
  \end{center}
\end{widetext}
%%%%%%%%%%%%%%%%%%%%%%%%%%%%%%%%%%%%%%%%%%%%%%%%%%%%%%%%%

\section{Symmetry aspect of anomalous Hall effect}
\label{Sec:Sym}
\subsection{Symmetry of the Berry curvature in ${\bm k}$ space}
\label{Sec:BerryCurv}
 The intrinsic AHC is expressed as the Berry curvature integrated over the
 Brillouin zone (BRZ) of one-electron bands below the Fermi
 level~\cite{Onoda2002,Jungwirth2002} such as: 
\begin{eqnarray}
\sigma_{\alpha\beta}=-\frac{e^2}{\hbar}\int
\frac{d{\bm k}}{(2\pi)^3}\sum_{n}f(\varepsilon_n({\bm
k})-\mu)\Omega_{n,\alpha\beta}({\bm k})\ ,
\label{Eq:sigma}
\end{eqnarray}
  where $n$ is the band index and $\alpha$, $\beta$=$x$, $y$, $z$ with
  $\alpha\ne\beta$.
  The Berry curvature for the AHC is defined as
\begin{eqnarray}
  \Omega_{n,\alpha\beta}({\bm k})=-2{\rm Im} \sum_{m\ne
 n}\frac{v_{nm,\alpha}({\bm k})v_{mn,\beta}({\bm
 k})}{[\varepsilon_{m}({\bm k})-\varepsilon_{n}({\bm k})]^2}
\label{Eq:omega}
\end{eqnarray}
 from the Kubo-formula~\cite{Thouless1982, Wang2006}.
 In these equations, the $\varepsilon_{n}({\bm k})$ is the eigenvalue
 and 
\begin{eqnarray}
v_{nm,\alpha}({\bm k})=\frac{1}{\hbar}\bigg\langle u_{n}({\bm k})\bigg| \frac{\partial \hat{H}({\bm k})}{\partial k_{\alpha}}
 \bigg| u_{m}({\bm k}) \bigg\rangle ,
\label{Eq:verocity}
\end{eqnarray}
 where $u_{n{\bm k}}$ is the periodic cell part of the Bloch states and
 $ \hat{H}({\bm k})=e^{-i{\bm k}\cdot{\bm r}}\hat{H}e^{i{\bm k}\cdot{\bm
 r}}$.
  For the convenience of our discussions, we hereafter use the vector form notations for
 the AHC and Berry curvature, i.e. ${\boldsymbol \sigma}=(\sigma^{x},
 \sigma^{y}, \sigma^{z}) \equiv (\sigma_{yz}, \sigma_{zx}, \sigma_{xy})$
 and ${\mathbf \Omega}_n= (\Omega_{n}^{x}, \Omega_{n}^{y},
 \Omega_{n}^{z}) \equiv (\Omega_{n,yz}, \Omega_{n,zx}, \Omega_{n,xy})$.

  From Eq. (\ref{Eq:sigma}), the appearance of the AHC,
  $\sigma^{\alpha}$, is governed by the Berry curvature in ${\bm k}$ space, $\Omega^{\alpha}({\bm k})$. 
  Thus let us first discuss the symmetry of the Berry curvature in ${\bm k}$ space.
  The group velocity is expressed with the Berry phase
  correction as follows~\cite{Chang1996,Sundaram1999}:
\begin{eqnarray}
\dot{{\bm
  r}}=\frac{1}{\hbar}\frac{\partial \varepsilon_{n}({\bm k})}{\partial
  {\bm k}}- \dot{{\bm k}}\times{\mathbf \Omega}_n({\bm k})\ .
 \label{Eq:groupvelocity}
\end{eqnarray}
   Transformation property of the Berry curvature with respect to the
   symmetry elements of magnetic space groups can be
   derived from this equation since the properties of
   $\varepsilon_{n}({\bm k})$, $\dot{{\bm r}}$, ${\bm k}$ and $\dot{{\bm
   k}}$ are known.
   First, the Berry curvature is not modified by any
   translation operations. Second, it is transformed as ordinary vectors for
   rotation operations in ${\bm k}$ space. Third, the space inversion
   brings ${\mathbf \Omega}_n({\bm k})$ to ${\mathbf \Omega}_n(-{\bm
   k})$. Thus the Berry curvature ${\mathbf \Omega}_n({\bm k})$
   behaves as an axial vector in ${\bm k}$ space. Finally, the time reversal
   operation transforms ${\mathbf \Omega}_n({\bm k})$ to $-{\mathbf \Omega}_n(-{\bm k})$.

    These transformation properties of the Berry curvature define
   constraints on its structure in ${\bm k}$ space.
   The well known relations are ${\mathbf \Omega}_n(-{\bm k})={\mathbf
   \Omega}_n({\bm k})$ for systems with the space inversion symmetry and ${\mathbf \Omega}_n(-{\bm k})=-{\mathbf
   \Omega}_n({\bm k})$ for systems with the time reversal symmetry. Some of other relations are listed in Table \ref{tab:BerryTransform}.
    These relations define further constraints on the Berry curvature at some
   ${\bm k}$ points, related to the elements of the group of ${\bm k}$.
    A simple example is that the Berry curvature is zero for all over the
   BRZ under both the space inversion and time reversal symmetries since the
   successive transformation of these operations results
   in ${\mathbf \Omega}_n({\bm k})=-{\mathbf \Omega}_n({\bm k})$,
   leading to ${\mathbf \Omega}_n({\bm k})=0$.
    Another example is the magnetic systems which have $T C_{2z}$
   symmetry. In this case, $\Omega^{z}({\bm k})=0$ on the $k_z=0$ plane
   since $\Omega^{z}(k_x,k_y,0)=-\Omega^{z}(k_x,k_y,0)$.
%%%%%%%%%%%%% Forbidden symmetry of AHC %%%%%%%%%%%%%%%
\begin{widetext}
  \begin{center}
  \begin{table}[tb]
    \caption{Complete list of symmetry operators and AHC components
   forbidden to be finite. The translation part of the operators,
   which does not affect the results, is not shown.
   In this table, all the superscripts of the AHC are explicitly
   written such as $\sigma^{\ell}_{ij}\equiv\sigma^{\ell}=\sigma_{ij}$
   ($i,j,\ell = x,y,z$).
   $C_{n(ij)}$ indicate the $n$-fold rotation operators whose rotation axes are in the
   ($ij$)-plane, $C_{n\mu}$ the $n$-fold rotation operators along the $\mu$-axis. 
   Mirror operator with the mirror plane normal to the $\mu$-axis
   is $PC_{2\mu}$. The operator replacing the
   $C_{n\mu}$ to $C_{n\mu}^{-1}$ also belongs to the same category in the list.
   The integer in parenthesis shows the number of the $O_{h}$ ($D_{6h}$)
   magnetic-point-group elements.}
   \begin{tabular}{ccc} \hline \hline
   \multicolumn{3}{c}{Cubic} \\ 
\hline
    AHC component     & Unitary & Anti unitary \\
 $\sigma^{k}_{ij}$ & $C_{n(ij)}$, $PC_{n(ij)}$ [n=2,4] (16) & $T
	    C_{nk}$, $TP C_{nk}$ [n=0,2,4] (16) \\
 $\sigma^{[111]}$ & $C_{2[1{\bar 1}0]}$, $C_{2[01{\bar 1}]}$,
	$C_{2[{\bar 1}01]}$,  & $T C_{n[111]}$, $T P C_{n[111]}$ [n=0,3] (6) \\
   &    $PC_{2[1{\bar 1}0]}$, $PC_{2[01{\bar 1}]}$,
	$PC_{2[{\bar 1}01]}$ (6)    & \\
\hline
   \multicolumn{3}{c}{Hexagonal} \\ 
\hline
    AHC component     & Unitary & Anti unitary \\
 $\sigma^{z}_{xy}$ & $C_{2(xy)}$, $PC_{2(xy)}$ (12) & $T
	    C_{nz}$, $T P C_{nz}$ [n=0,2,3,6] (12) \\
 $\sigma^{x}_{yz}$ & $C_{nz}$, $PC_{nz}$ [n=2,3,6]
 & $T C_{nz}$, $T P C_{nz}$ [n=0,3,6] \\
& $C_{2y}$, $PC_{2y}$ (12) & $T C_{2x}$,
	    $T P C_{2x}$ (12) \\
 $\sigma^{y}_{zx}$ & $C_{nz}$, $PC_{nz}$ [n=2,3,6]
 & $T C_{nz}$, $T P C_{nz}$ [n=0,3,6] \\
& $C_{2x}$, $PC_{2x}$ (12) & $T C_{2y}$,$T P C_{2y}$ (12) \\
\hline \hline
   \end{tabular}
\label{tab:ForbidSym}
  \end{table}
  \end{center}
\end{widetext}
%%%%%%%%%%%%%%%%%%%%%%%%%%%%%%%%%%%%%%%%%%%

\subsection{Symmetry condition of finite AHC}
\label{Sec:SymAHC}
%%%%%%%%%%%%% Transformation property of Berry curvature %%%%%%%%%%%%%%%
\begin{figure}[tb]
	\includegraphics[width=1.0\linewidth]{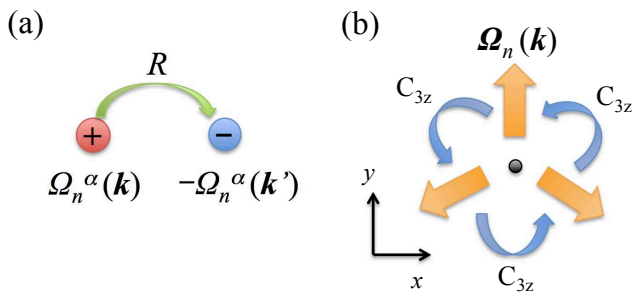}
	\caption{Transformation of the Berry curvature under the symmetry operators. (a) Operators which
 reverse the sign of the Berry curvature and (b) the three-fold
 rotation along the $z$-axis.} 
	\label{Fig:Berry_Trans}
\end{figure}
%%%%%%%%%%%%%%%%%%%%%%%%%%%%%%%%%%%%%%%%%%%%%%%%%%%%%%%%%%%%%%%%%%%%%%%
   From Eq.\ (\ref{Eq:sigma}), the symmetry of the Berry
 curvature in ${\bm k}$ space, discussed in the previous subsection,
 determines whether
 $\sigma^{\alpha}$ can be finite or not in a magnetic system.
Since the Berry curvature is not affected by any translation symmetries,
it is enough to take only the magnetic point group into account.
  If the Berry curvature satisfies the condition $\Omega^{\alpha}(R{\bm
 k})=-\Omega^{\alpha}({\bm k})$ due to a magnetic symmetry, the corresponding AHC
component $\sigma^{\alpha}$ must be zero since the Berry
curvature at ${\bm k}$ and $R{\bm k}$ are canceled out by the BRZ integration (see
Fig. \ref{Fig:Berry_Trans} (a)). In this case, $\Omega^{\alpha}({\bm
 k})$ is zero when $R$ is an element of the group of ${\bm
 k}$, i.e. $R{\bm k}={\bm k}$.
  Similarly, when a magnetic system has an $n$-fold rotation symmetry, the Berry
 curvature is canceled out through the BRZ integration in Eq.\ (\ref{Eq:sigma}) (Fig. \ref{Fig:Berry_Trans} (b)).
 The components of ${\boldsymbol \sigma}$ normal to the $n$-fold axis thus
 disappear. For example, $\sigma^{x}$ and $\sigma^{y}$ are zero when the
 system has a rotation symmetry with respect to the $z$-axis.
 We provide a complete list of the relation
  between the symmetry operators and the forbidden components
of the AHC in Table \ref{tab:ForbidSym}.
 The AHC component can be finite when the magnetic order
 breaks all of the corresponding symmetries listed in Table
 \ref{tab:ForbidSym}. 
  Structures of the AHC tensors under all the magnetic point-group
 symmetries have been listed in
 Refs. \onlinecite{Kleiner1966,Seemann2015} by considering the
transformation coefficients for the operators of magnetic point
 groups. Table \ref{tab:ForbidSym} is equivalent to the lists in
 these previous works.
  Note that the symmetry operators in Table \ref{tab:ForbidSym}
 also forbid finite magnetization $M_{\alpha}$. This is because
 $M_{\alpha}$ and $\sigma^{\alpha}$ have the same
 transformation property for the magnetic symmetry operations, which is a
 natural consequence from the same transformation property of the Berry
 curvature and that of the magnetic moment in ${\bm k}$ space with
 respect to the operators of the magnetic space group.
 
\subsection{S-O coupling and AHC}
\label{Sec:SO}
%%%%%%%%%%%%% Global spin rotation  %%%%%%%%%%%%%%%
\begin{figure}[tb]
	\includegraphics[width=1.0\linewidth]{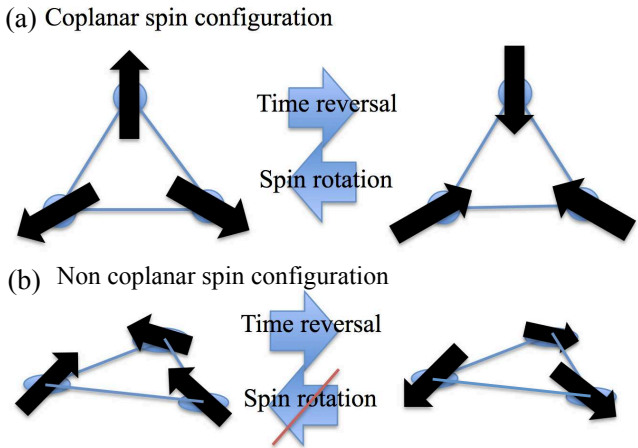}
	\caption{Time reversal operation and global spin rotation for a (a) coplanar
 spin configuration and (b) non coplanar spin configuration in
 the triangular system.} 
	\label{Fig:SpinConfig_TimeRev}
\end{figure}
%%%%%%%%%%%%%%%%%%%%%%%%%%%%%%%%%%%%%%%%%%%%%%%%%%
  The effect of the S-O coupling on the AHE has been one of the fundamental issues
  since the pioneering work by Karplus and Luttinger~\cite{Karplus1954}. 
  Here, we provide a comprehensive explanation for the relation between
  the S-O coupling and the AHE in general magnetic states.
   First, let us note that the symmetry group for a nonmagnetic system
  without the S-O coupling is expressed as $\mathcal{M}_{\rm
  nmag,nso}=\mathcal{M}_{\rm para}\times SU(2)$, where $\mathcal{M}_{\rm
  para}=\mathcal{G}\times\{E,T\}$, $\mathcal{G}$ is the ordinary space
  group of the crystal structure and $E$ is the identity element
  of the space group.
    Magnetic order breaks both $\mathcal{M}_{\rm para}$ and
  $SU(2)$. However, in general, the symmetry of a magnetic system without
  the S-O coupling is higher than that with the S-O coupling by
  the following reason.
    The symmetry group of a magnetic system without the S-O coupling,
  $\mathcal{M}_{\rm mag,nso}$, belongs to a subgroup of
  $\mathcal{M}_{\rm nmag,nso}$, and that with the S-O coupling,
  $\mathcal{M}_{\rm mag,so}$, belongs to a subgroup of $\mathcal{M}_{\rm
  para}$ because the S-O coupling breaks all of the symmetries
  related to $SU$(2). Therefore, $\mathcal{M}_{\rm mag,so}$ is a subgroup of
  $\mathcal{M}_{\rm mag,nso}$. 
   As discussed in section \ref{Sec:BerryCurv}, the
  transformation property of the Berry
  curvature with respect to magnetic symmetry operations is similar to
  that of spin in ${\bm k}$ space, i.e. it is transformed as an axial
  vector and reversed by $T$.
  Meanwhile, the spin rotation $R_{s}(\theta_S,\phi_S)$ does not affect
  the Berry curvature in the absence of the S-O coupling.
  As a result, the magnetic symmetries listed in Table
  \ref{tab:ForbidSym} preserved by further multiplying the spin
  rotations also forbid the corresponding $\sigma^{\alpha}$ to be
  finite in magnetic systems without the S-O coupling.

 Ordinary collinear FM system is the most fundamental example in which
 the S-O coupling is required to induce the AHE. 
 The FM systems without the S-O coupling always preserve the time reversal
 symmetry with the spin rotations $R_{S}(\theta_S, \phi_S)$.
 The $R_{S}T$ symmetry preserved in the system without the S-O coupling was referred as
 ``effective $T$ symmetry'' in Ref. \onlinecite{Gosalbez2015}. 
   Coplanar AFM spin configurations also require the S-O coupling
 to induce the AHE due to the $R_{S}T$ symmetry preserved in the
 absence of the S-O coupling, since
 $T$ works as the 180-degrees-spin rotation around the
 axis normal to the coplanar plane [Fig.\ \ref{Fig:SpinConfig_TimeRev} (a)].
  If the spin moments rise up from the coplanar plane, the spin configuration
 after operating $T$ can not be brought back with
the spin rotation $R_{S}$ due to the spin components normal to the
 coplanar plane [Fig.\ \ref{Fig:SpinConfig_TimeRev} (b)].  In
 this case, the non-coplanar spin system breaks the $R_{S}T$
 symmetry as well as $T$ symmetry
 and the AHC therefore can be finite without the S-O coupling.
  This idea also explains why scalar spin chirality can induce the
 AHE~\cite{Shindou2001}, since finite scalar spin chirality
 always breaks the $R_{S}T$ symmetry.

   Mn$_3$Ir, Mn$_3$Sn, and Mn$_3$Ge undergo coplanar magnetic
 order and require the S-O coupling to induce the AHE from the above
 discussions. Indeed, the first-principles calculation for Mn$_3$Ir
 confirmed that a finite spin component normal to the coplanar plane is
 required for finite AHC in the absence of the S-O coupling~\cite{Chen2014}.

\section{Cluster multipole moments in AFM states}
\label{Sec:Multipole}
\subsection{Definition of CMP}
\label{Sec:DefCMP}
     As discussed in Sec.\ \ref{Sec:SymAHC}, magnetic structures
    induce finite AHC $\sigma^{\alpha}$ in the same symmetry
    condition for finite magnetization $M_{\alpha}$ in the presence of the S-O coupling.
    This means that the magnetic systems having net
    magnetization always belong to the symmetry which can induce
      the AHE with the S-O coupling.
     On the other hand, finite magnetization is not necessary to induce
    the AHE in AFM states. The question is then ``is there a
    macroscopic order parameter which characterizes the AHE?''.
      For identifying such an order parameter in generic magnetic
    states, we here introduce a quantity, which we call
     a cluster multipole moment (CMP).
     With the framework based on the CMP, we can quantify the symmetry
    breaking due to commensurate non-collinear magnetic order.
     Below, we provide general theory of the CMP. A concrete example is provided in Sec. \ref{Sec:CMP_Mn3Sn} for
Mn$_3$Ir and Mn$_3$$Z$.

  We first identify atom clusters for which we define the CMP moments.
  In general, a crystal contains a number of atoms which are inequivalent
under the crystal symmetry.
 Each atom cluster is defined as a set of atoms related with one another by the {\it crystal} symmetry 
operators without space translation in the {\it magnetic} unit cell.
For simplicity, we here consider only the case of the magnetic order
characterized by the wave vector ${\bm q}={\bm 0}$, whose magnetic unit
  cell is the same with that of the crystal unit cell~\cite{Note2}.
   A space group $\mathcal{G}$, which describes the symmetry of
  a crystal structure, is decomposed into the cosets of the 
  maximum symmorphic subgroup ${\mathcal  H}$ as 
\begin{eqnarray}
 {\mathcal G} = \sum_{i=1}^{N_{\rm coset}}\{R_{i}|{\boldsymbol \tau}_{i}\}{\mathcal H}\ ,
\label{Eq:coset} 
\end{eqnarray}
 where $N_{\rm coset}$ is the number of the cosets, $R_i$ and ${\boldsymbol
 \tau}_{i}$ represent the point-group element and translation
 operator of the element in the space group $\mathcal{G}$,
 respectively, with $\{R_{1}|{\boldsymbol
 \tau}_{1}\}\equiv\{E|\mathbf{0}\}$ and ${\boldsymbol
 \tau}_{i}\neq \mathbf{0}$ for $i\geq 2$.
  In Eq. (\ref{Eq:coset}), $N_{\rm coset}=1$ for crystal
 structures which belong to symmorphic space groups and
 $N_{\rm coset}> 1$ for crystal structures which belong to
 nonsymmorphic space groups. Therefore, nonsymmorphic crystal structures
 contain multiple clusters related with one another by the symmetry
 operators $\{R_{i}|{\boldsymbol \tau}_{i}\}$ in the unit cell.
 The origin of the cluster is naturally defined as the point
 which satisfies all the point symmetries for which the cluster is defined.

 Analogous to the local multipole moments defined for
 an atom~\cite{Kusunose2008,Kuramoto2008,Santini2009}, here we define
 the rank-$p$ CMP moment for the $\mu$-th cluster as
follows:
\begin{eqnarray}
      M_{pq}^{(\mu)}\equiv
       \sqrt{\frac{4\pi}{2p+1}}\sum_{i=1}^{N^{(\mu)}_{\rm atom}}{\bm
       m}_{i}\cdot \nabla_{i} (|{\bm
       R}_{i}|^{p}Y_{pq}(\theta_i,\phi_i)^{*})\ ,
\label{Eq:multipole}
\end{eqnarray}
 where $N^{(\mu)}_{\rm atom}$ is the number of atoms of the $\mu$-th
cluster, ${\bm m}_{i}$ is a magnetic moment on the $i$-th
atom, $\nabla_{i}\equiv\frac{\partial}{\partial {\bm R}_{i}}$, ${\bm
R}_i\equiv(X_i,Y_i,Z_i)$ is the position of the $i$-th atom, $Y_{pq}$
are the spherical harmonics, and $R_{i}$, $\theta_i$ and $\phi_i$ are the
distance, polar angle and azimuthal angle, respectively, of
the $i$-th atom. 
Based on the Wannier bases, $\{w_{i,a}\}$, the magnetic moment on the $i$-th atom is calculated
as follows:
\begin{widetext}
 \begin{eqnarray}
  {\bm m}_{i}&=& \mu_{B}\sum_{n}\sum_{ab} \int\frac{d{\bm
  k}}{(2\pi)^3} f(\varepsilon_{n}({\bm k})-\mu) \nonumber \\ 
  &\times& \langle u_{n}({\bm
  k}) | w_{i,a} \rangle \langle w_{i,a} | ({\bm \ell}+2{\bm s})|
  w_{i,b} \rangle \langle w_{i,b} |u_{n}({\bf
  k})\rangle\ ,
\label{Eq:magmom}
\end{eqnarray}
\end{widetext}
where $\mu_{B}=-|e|\hbar/2m$ is the Bohr magneton, and ${\bm \ell}$ and
${\bm s}$ are the orbital and spin angular momentum operators.
%  Specific magnetic clusters can be characterized with a magnetic
%  monopole~\cite{Castelnovo2008,Khomskii2012}. The magnetic monopole can
%  be defined using the analogy to the integral of the magnetic charge
%  density~\cite{Kusunose2008}, such as:
%\begin{eqnarray}
% M_{00}\equiv\frac{1}{\sqrt{4\pi}}\sum_{i=1}^{N_{\rm atom}}\sum_{\alpha=X,Y,Z} M_{i\alpha_i}/\alpha_i
%\end{eqnarray}
The macroscopic contribution of the CMP moment can be defined by the summation
over the clusters in the magnetic unit cell:
\begin{eqnarray}
      M_{pq}=\frac{N^{\rm u}_{\rm atom}}{N^{\rm c}_{\rm atom}}\frac{1}{V}\sum_{\mu=1}^{N_{\rm cluster}}M_{pq}^{(\mu)}\ ,
\label{Eq:MPsum}
\end{eqnarray}
where $V$ is the volume of the magnetic unit cell, $N^{\rm u}_{\rm atom}$ is
the number of atoms in the magnetic unit cell, $N^{\rm c}_{\rm
atom}=\sum_{\mu}N^{(\mu)}_{\rm atom}$ is the total number of atoms in all
of the clusters, and $N_{\rm cluster}$ is the number of clusters in the
unit cell, which is the same with the multiplication of $N_{\rm coset}$
and the number of atoms inequivalent under the space-group symmetry.

\subsection{Symmetry classification of CMP moments}
\label{Sec:CMP_Symmetry}
\begin{figure}[t]
	\includegraphics[width=1.0\linewidth]{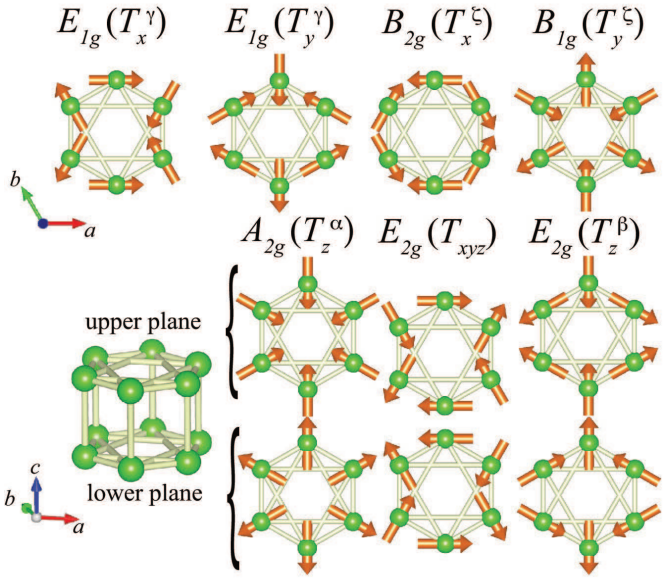}
	\caption{AFM structures characterized by the cluster octupole
 moments of the $D_{6h}$ IREPs in Table \ref{tab:HexMP}.} 
	\label{Fig:MagnCMPs}
\end{figure}

    The local multipole moments for $f$-electron systems are
   classified according to the irreducible representations (IREPs) of
   the point group symmetry of the atomic site~\cite{Shiina1997,Kiss2005,Takimoto2006,Kusunose2008,Kuramoto2009,Santini2009,Suzuki2014}.
    Similarly, the CMP moments can be classified according to the point group
   symmetry of the atomic configuration.
For a crystal structure whose conventional crystal axes are
orthogonal, the CMP moments are classified according
to the IREPs of the $O_{h}$ point group in Table \ref{tab:CubMP}.
For a crystal structure with the hexagonal conventional axes
such as the hexagonal and trigonal lattice systems, the
  CMP moments classified according to the $D_{6h}$ IREPs should be
  used to reflect the point group symmetry. In Table \ref{tab:HexMP}, we
   provide a list of the $D_{6h}$ CMP moments.
 These CMP moments are all odd with respect to the time reversal operator $T$. If the
 magnetic structure preserves the inversion symmetry, only odd rank
 CMP moments can be finite due to the relation
   $M_{pq}=(-)^{p+1}M_{pq}$ in Eq. (\ref{Eq:multipole}). Figure
   \ref{Fig:MagnCMPs} shows non-collinear magnetic structures
   characterized by the lowest rank cluster octupole moments for the
   $D_{6h}$ IREPs.

\begin{widetext}
\begin{center}
 \begin{table}[tb]
  \caption{ CMP moments up to rank three
  classified according to the IREPs of the $O_{h}$ point group.
    The quadrupole CMP moments can be finite only 
  for magnetic structures without the space inversion symmetry (see text).
   Note that this table is analogous
  to the {\it magnetic} multipole moments classified according to the
  $O_{h}$ point group, and the similar list is provided for electric multipole
  moments in Ref. \onlinecite{Shiina1997}.}
  \begin{tabular}{ccc}
\hline \hline
                     &  IREP   &           CMP         \\
\hline
%{\blue Monopole}    & $A_{1u}$ &  $N_{0}\equiv M_{00}$ \\
%\hline
    Rank 1          & $T_{1g}$ & $J_{x}\equiv \frac{1}{\sqrt{2}}(-M_{11}+M_{1-1})$  \\
   (Dipole)         &                    & $J_{y}\equiv -\frac{i}{\sqrt{2}}(M_{11}+M_{1-1})$   \\
                    &                    & $J_{z}\equiv M_{10}$\\
\hline
    Rank 2         & $E_{u}$ & $Q_{3z^2-r^2}\equiv M_{20}$ \\
  (Quadrupole)            &                 &
	   $Q_{x^2-y^2} \equiv \frac{1}{\sqrt{2}}(M_{22}+M_{2-2})$ \\
                        & $T_{2u}$ &
	   $Q_{yz} \equiv -\frac{i}{\sqrt{2}}(M_{21}+M_{2-1})$ \\
                        &       &
	   $Q_{zx} \equiv \frac{1}{\sqrt{2}}(-M_{21}+M_{2-1})$ \\
                        &       &
	   $Q_{xy} \equiv \frac{i}{\sqrt{2}}( M_{22}-M_{2-2})$ \\
\hline
   Rank 3       & $A_{2g}$ &
	   $T_{xyz} \equiv \frac{i}{\sqrt{2}}( M_{32}-M_{3-2})$ \\
  (Octupole)         & $T_{1g}$ &  
	   $T^{\alpha}_{x} \equiv \frac{1}{4}[\sqrt{5}(-M_{33}+M_{3-3})-\sqrt{3}(-M_{31}+M_{3-1}) ]$  \\
                      &                   &
	   $T^{\alpha}_{y} \equiv \frac{i}{4}[\sqrt{5}(M_{33}+M_{3-3})+\sqrt{3}(M_{31}+M_{3-1})]$ \\
                      &                   &
	   $T_z^{\alpha} \equiv M_{30}$ \\
                      & $T_{2g}$ &
	   $T^{\beta}_{x} \equiv -\frac{1}{4}[\sqrt{3} (-M_{33}+M_{3-3})+\sqrt{5}(-M_{31}+M_{3-1})]$ \\
                      &                      &
	   $T^{\beta}_{y} \equiv \frac{i}{4}[ \sqrt{3}(M_{33}+M_{3-3})-\sqrt{5}(M_{31}+M_{3-1})$ \\
                      &       &
	   $T^{\beta}_{z} \equiv \frac{1}{\sqrt{2}}(M_{32}+M_{3-2})$ \\
\hline \hline
  \end{tabular}
\label{tab:CubMP}
% \end{table}
%\end{center}
%\end{widetext}
%\begin{widetext}
%\begin{center}
% \begin{table}[h]
\caption{ CMP moments up to rank three classified according to the IREPs
  of the $D_{6h}$ point group. Note that this table is analogous
  to the {\it magnetic} multipole moments classified according to the
  $D_{6h}$ point group. The quadrupole CMP moments can be finite only
  for magnetic structures without the space inversion symmetry (see text).}
  \begin{tabular}{ccc}
\hline \hline
                     &  IREP   &           CMP         \\
\hline
%{\blue Monopole}     & $A_{1u}$ &  $N_{0}\equiv M_{00}$ \\
%\hline
   Rank 1          & $A_{2g}$ & $J_{z} \equiv M_{10}$ \\
  (Dipole)               & $E_{1g}$ & $J_{x} \equiv \frac{1}{\sqrt{2}}(-M_{11}+M_{1-1})$ \\
                    &                  & $J_{y} \equiv -\frac{i}{\sqrt{2}}(M_{11}+M_{1-1})$  \\
\hline
     Rank 2         & $A_{1u}$ & $Q_{3z^2-r^2} \equiv M_{20}$ \\
   (Quadrupole)                   & $E_{2u}$ &
	   $Q_{x^2-y^2} \equiv \frac{1}{\sqrt{2}}(M_{22}+M_{2-2})$ \\
                        &                      &
	   $Q_{xy} \equiv \frac{i}{\sqrt{2}}( M_{22}-M_{2-2})$ \\
                        & $E_{1u}$ &
	   $Q_{zx} \equiv \frac{1}{\sqrt{2}}(-M_{21}+M_{2-1})$ \\
                        &       &
	   $Q_{yz} \equiv -\frac{i}{\sqrt{2}}(M_{21}+M_{2-1})$ \\
\hline
    Rank 3  & $A_{2g}$ &  $T_z^{\alpha} \equiv M_{30}$ \\
 (Octupole) & $E_{1g}$ &
	   $ T^{\gamma}_{x} \equiv \frac{1}{\sqrt{2}}(-M_{31}+M_{3-1})$ \\
                      &       &
	   $ T^{\gamma}_{y} \equiv -\frac{i}{\sqrt{2}}(M_{31}+M_{3-1})$ \\
                      & $E_{2g}$ &
	   $ T_{xyz} \equiv \frac{i}{\sqrt{2}}( M_{32}-M_{3-2})$ \\
                      &       &
	   $ T^{\beta}_{z} \equiv \frac{1}{\sqrt{2}}( M_{32}+M_{3-2})$ \\
                      & $B_{2g}$ &
	   $ T^{\zeta}_{x} \equiv \frac{1}{\sqrt{2}}(-M_{33}+M_{3-3})$ \\
                      & $B_{1g}$      &
	   $ T^{\zeta}_{y} \equiv \frac{i}{\sqrt{2}}(M_{33}+M_{3-3})$ \\
\hline \hline
  \end{tabular}
\label{tab:HexMP}
 \end{table}
\end{center}
\end{widetext}

%%%%%%%%%%%%% Examples of AFM states of D6h CMP moments %%%%%%%%%%%%%%%
%\begin{widetext}
%\begin{center}
%\end{center}
%\end{widetext}
%%%%%%%%%%%%%%%%%%%%%%%%%%%%%%%%%%%%%%%%%%%%%%%%%%%%%%%%%%%%%%%%%%%%%%
%%%%%%%%%%%%% Crystal structure and CMP magnetic structure of Mn3Ir %%%%%%%%%%%%%%%
\begin{figure}[tb]
	\includegraphics[width=1.0\linewidth]{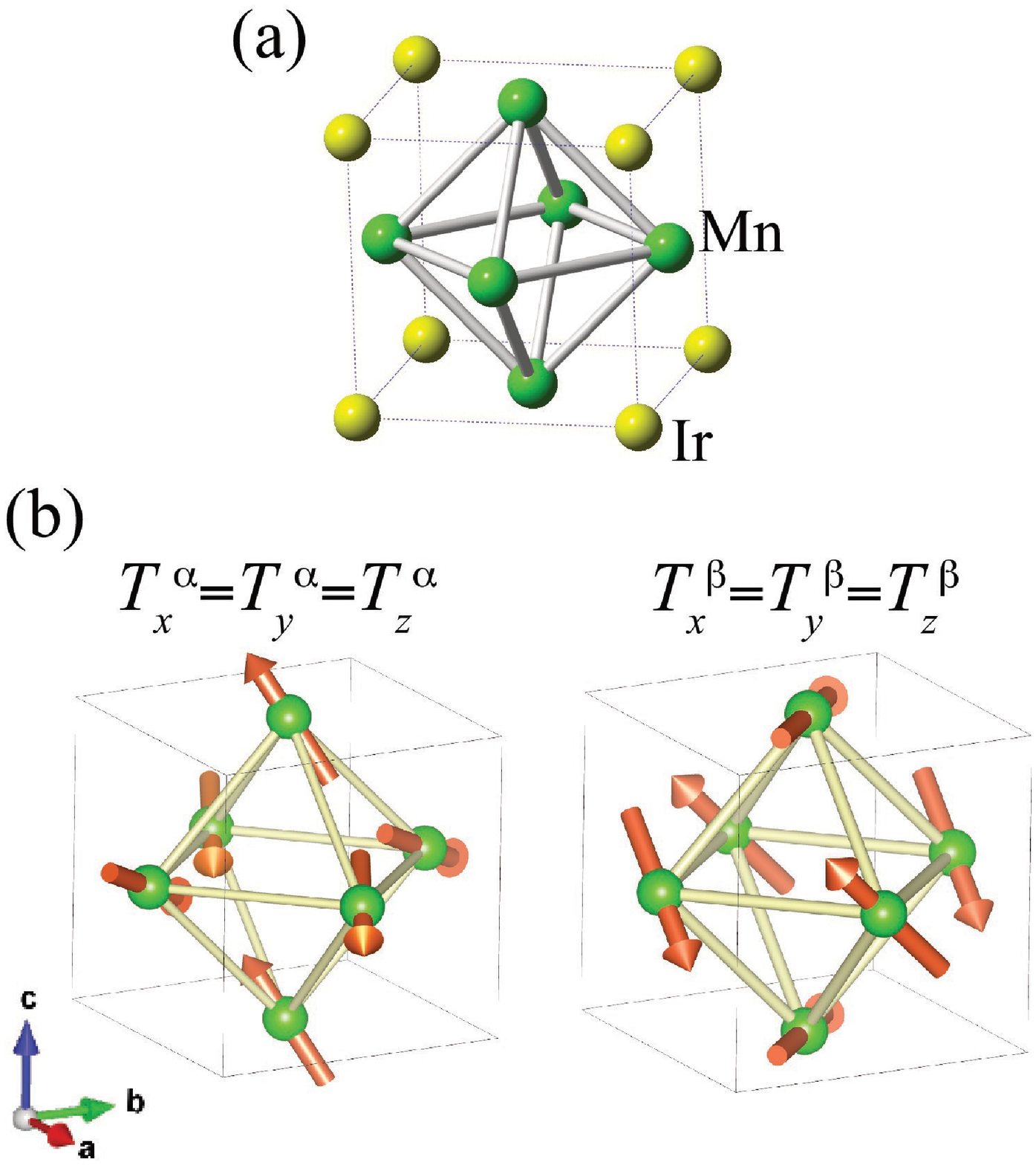}
	\caption{ (a) Crystal structure of Mn$_3$Ir. (b)
 The AFM structures characterized by the cluster octupole moments of 
 $T_{1g}$ CMP with the relation $T_{x}^{\alpha}$=
 $T_{y}^{\alpha}$=$T_{z}^{\alpha}$, which has
 been recognized experimentally, and of $T_{2g}$ CMP with $T_{x}^{\beta}$=$T_{y}^{\beta}$=$T_{z}^{\beta}$
 in the Mn cluster of Mn$_3$Ir.
} 
	\label{Fig:Crystal_Mn3Ir}
\end{figure}
%%%%%%%%%%%%%%%%%%%%%%%%%%%%%%%%%%%%%%%%%%%%%%%%%%%%%%%%%%%%%%%%%%%%%%
Ordinary magnetization $M_{\alpha}$ ($\alpha=x,y,z$) corresponds
to the macroscopic contribution of the cluster dipole moment $J_{\alpha}$.
 Meanwhile, magnetization of non-collinear AFM states without
 net dipole magnetization is characterized by the macroscopic
 contribution of CMP moments with the ranks higher than one.
  From the discussion of Sec.\ \ref{Sec:SymAHC},
 $\sigma^{\alpha}$, $M_{\alpha}$, and $J_{\alpha}$ are transformed in
 the same manner for the operation of the magnetic-point-group
 elements, which means that they belong to the same IREPs of the $O_{h}$ and
 $D_{6h}$ point groups.
 The conditions for the AHE can now be concisely described with the
 symmetrized CMP moments. The AHE is induced with the emergence
 of the finite magnetization of the CMP moments which belong to
 the same IREP of dipole moments, i.e. $T_{1g}$ ($A_{2g}$ and/or
 $E_{1g}$) CMP moments in the $O_{h} (D_{6h})$ representation.
  Note that, in the absence of the S-O coupling, the AHE requires
 $R_{s}T$-symmetry breaking as well as magnetization of these CMP moments
 as discussed in Sec.\ \ref{Sec:SO}.

\subsection{CMP moment and AHE in Mn$_3$Ir and Mn$_3$$Z$}
\label{Sec:CMP_Mn3Sn}
   Let us now apply the scheme discussed above to the AFM
   spin configurations observed in Mn$_3$Ir and Mn$_3$$Z$, for which the
   AHE in the AFM states has been studied~\cite{Chen2014, Nakatsuji2015,
  Kiyohara2016,Nayak2016}.
    Mn$_3$Ir crystallizes into the simple cubic structure which belongs to the
   space group $Pm\overline{3}m$ ($O_{h}^{1}$, No. 221) as shown in 
   Fig. \ref{Fig:Crystal_Mn3Ir}. This crystal structure leads to $N_{\rm
   coset}=1$ in Eq. (\ref{Eq:coset}) since the space group is symmorphic.
    Here, we focus on the Mn atoms, which have the finite magnetic moments
   in the AFM state.
     The unit cell contains three Mn atoms ($N_{\rm atom}^{u}$=3), and 
   we can define a Mn cluster, which contain six Mn atoms ($N_{\rm atom}^{c}$=6) related to
   each other with the operation of the symmetry elements of the $O_h$
   point group.
    In Fig. \ref{Fig:Crystal_Mn3Ir} (b), we show the AFM structures
   characterized by $T_{1g}$ octupole moment with the order parameter
   $T_{x}^{\alpha}$=$T_{y}^{\alpha}$=$T_{z}^{\alpha}$, which has been recognized experimentally, 
  and $T_{2g}$ octupole moment with $T_{x}^{\alpha}$=$T_{y}^{\alpha}$=$T_{z}^{\alpha}$.
   Following the discussion in
   Sec. \ref{Sec:CMP_Symmetry}, the AFM configuration with $T_{1g}$
   cluster octupole moments, which belong to the same IREP of the dipole moment, 
   can induce the AHE, as predicted in the earlier work by first-principles
   calculation~\cite{Chen2014}, while that with $T_{2g}$ does not.
    The crystal symmetries broken by the $T_{1g}$ cluster octupole
   moment are
   completely the same as those broken by the co-linear magnetic dipole
   order along a [111] direction. As a result, the magnetic space group
   ($R\overline{3}m'$) for the AFM state with
   the $T_{1g}$ CMP, whose symmetry elements are listed in Table
   \ref{tab:MagSysSym}, is in common with the FM state.

\begin{figure}[tb]
	\includegraphics[width=1.0\linewidth]{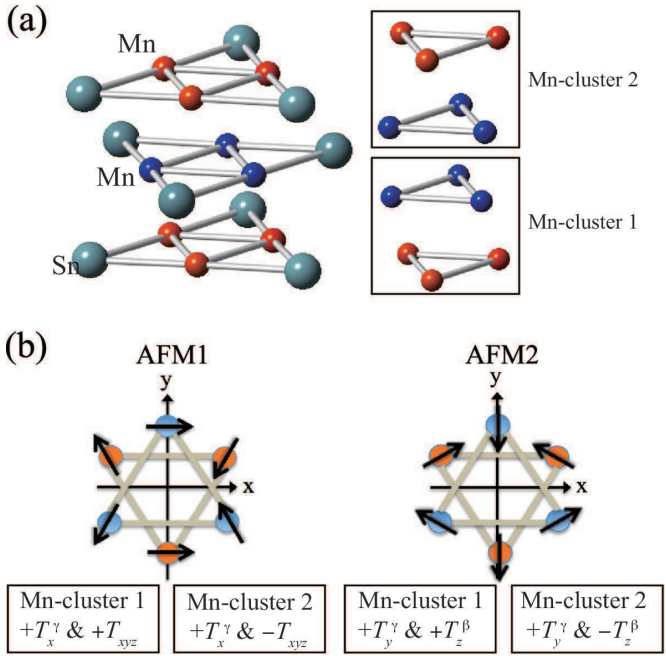}
	\caption{ (a) Crystal structure of Mn$_3$$Z$ ($Z$=Sn, Ge) with
 the Mn-clusters defined by the space group (see text). (b)
 Spin configuration on the Mn atoms in Mn$_3$Sn. The AFM1 and AFM2
 spin structures are experimentally realized depending on the direction of magnetic
 fields along the $x$ and $y$ directions,
 respectively~\cite{Tomiyoshi1982}.
The lowest rank CMP moments characterizing the spin configurations are also
 shown for each Mn-cluster.} 
	\label{Fig:Crystal_Mn3Sn}
\end{figure}

     Mn$_3$$Z$ crystallizes into the Ni$_3$Sn-type structure as shown
  in Fig. \ref{Fig:Crystal_Mn3Sn} (a).
   The hexagonal structure of Mn$_3$$Z$ belongs to the space group
  $P6_3/mmc$ ($D_{6h}^4$, space group No. 194)~\cite{Tomiyoshi1982a,Brown1990}.
  The nonsymmorphic space group $P6_3/mmc$ is
  decomposed into the cosets of the symmorphic space
  group $P\bar{3}m1$ ($D_{3d}^3$, No. 164) as $P6_3/mmc =
  P\bar{3}m1+\{C_{2z}| {\boldsymbol \tau}\}P\bar{3}m1$
  ($D_{3d}^3+\{C_{2z}| {\boldsymbol \tau}\}D_{3d}^3$), where
  ${\boldsymbol \tau}=(0,0,c/2)$. Following the discussion of
  Sec. \ref{Sec:Multipole}, Mn$_3$$Z$ contains four clusters, i.e. two
  clusters related to each other with the operation of
  $\{C_{2z}| {\boldsymbol \tau} \}$ for Mn and $Z$ atoms in
  the magnetic unit cell (see Fig. \ref{Fig:Crystal_Mn3Sn} (a)).
  Each cluster has the point group symmetry of $D_{3d}$.
  Since the magnetic moments in $Z$ atoms are negligible, we here ignore
  the $Z$-atom clusters. The macroscopic
  contribution of the CMP moment is calculated from Eq. (\ref{Eq:MPsum}) with
  $N^{\rm u}_{\rm
  atom}/N^{\rm c}_{\rm atom}=1/2$, since the unit cell of Mn$_3$$Z$
  contains six Mn-atoms ($N^{\rm u}_{\rm
  atom}=6$) and two Mn clusters consist of six Mn atoms for
  each cluster ($N^{\rm c}_{\rm atom}=12$) as shown in Fig. \ref{Fig:Crystal_Mn3Sn} (a).
   The $D_{3d}$ point group has six IREPs
  $A'_{1g/u}$, $A'_{2g/u}$, $E'_{g/u}$ in which the prime is for
  distinguishing the IREPs from those of the $D_{6h}$ point group. 
  The compatibility relations between the $D_{3d}$ and $D_{6h}$ IREPs
  are as follows:
\begin{eqnarray}
A_{1g/u}\downarrow D_{3d} = B_{2g/u}\downarrow D_{3d} = A'_{1g/u}  \nonumber \\
A_{2g/u}\downarrow D_{3d} = B_{1g/u}\downarrow D_{3d} = A'_{2g/u}  \nonumber \\
E_{1g/u}\downarrow D_{3d} = E_{2g/u}\downarrow D_{3d} = E'_{g/u} \ . \nonumber
\end{eqnarray}
  These relations mean, for instance, $A_{1g}$ and $B_{2g}$ CMP moments belong to
  the same IREP, $A'_{1g}$, in the cluster with the $D_{3d}$ point group
  symmetry.
%%%%%%%%%%%%% Crystal and magnetic structures of Mn3Sn %%%%%%%%%%%%%%%
%%%%%%%%%%%%%%%%%%%%%%%%%%%%%%%%%%%%%%%%%%%%%%%%%%%%%%%%%%%%%%%%%%%%%%
%%%%%%%%%%%%   Magnetic symmetry of Mn3Zn  %%%%%%%%%%%%%%%%%%%%%%%%%%%
\begin{center}
  \begin{table}[tb]
    \caption{Magnetic symmetry operators preserved in the AFM states of
   Mn$_3$Ir and Mn$_3$Sn.  
    In Mn$_3$Ir, the directions
   of the rotation axes are expressed with the cubic (Cartesian) axis.
   In Mn$_3$Sn, the $x$,$y$ axes are shown in Fig. \ref{Fig:Crystal_Mn3Sn} (b) and the $z$ axis is the normal direction of
   the $xy$ plane. ${\boldsymbol \tau}$ represents the translation (0,
   0, c/2). }
   \begin{tabular}{cc} 
\hline \hline
     Spin struct. &  Preserved Symmetries \\
\hline
           Mn$_3$Ir      &  $\{E\mid \mathbf{0}\},
	\{C_{3[111]}^{+}\mid\mathbf{0}\},\{C_{3[111]}^{-}\mid
	\mathbf{0}\}$, \\
                   & $T \{C_{2[1{\bar 1}0]}\mid \mathbf{0}\}, T
	\{C_{2[01{\bar 1}]}\mid \mathbf{0}\}, T \{C_{2[{\bar 1}01 ]}\mid
	\mathbf{0}\}$ \\
                   &  $\{P\mid \mathbf{0}\},
	\{PC_{3[111]}^{+}\mid\mathbf{0}\},\{PC_{3[111]}^{-}\mid
	\mathbf{0}\}$, \\
                   &  $T \{PC_{2[1{\bar 1}0]}\mid \mathbf{0}\}, T
	\{PC_{2[01{\bar 1}]}\mid \mathbf{0}\}, T \{PC_{2[{\bar 1}01 ]}\mid
	\mathbf{0}\}$ \\
     Mn$_3$Sn AFM1 & $\{E\mid \mathbf{0}\}, \{C_{2x}\mid\mathbf{0} \}, T
	\{C_{2z}\mid {\boldsymbol \tau}\}, T\{C_{2y}\mid {\boldsymbol
	\tau}\}$, \\
               & $\{P\mid \mathbf{0}\}, \{PC_{2x}\mid\mathbf{0} \}, T
	\{PC_{2z}\mid {\boldsymbol \tau}\}, T\{PC_{2y}\mid {\boldsymbol
	\tau}\}$ \\

     Mn$_3$Sn AFM2 & $\{E\mid \mathbf{0}\}, \{C_{2y}\mid {\boldsymbol
	\tau}\},T \{C_{2z}\mid {\boldsymbol \tau}\},
	T\{C_{2x}\mid\mathbf{0} \}, $\\
               & $\{P\mid \mathbf{0}\}, \{PC_{2y}\mid {\boldsymbol
	\tau}\}, T\{PC_{2z}\mid {\boldsymbol \tau}\}, T\{PC_{2x}\mid\mathbf{0} \} $\\
     \hline \hline
   \end{tabular}
\label{tab:MagSysSym}
  \end{table}
\end{center}
%%%%%%%%%%%%%%%%%%%%%%%%%%%%%%%%%%%%%%%%%%%%%%%%%%%%%%%%%%%%%%%%%%%%%%
%%%%%%%%%%%% Local moments vs octupole moments in AFM1 of Mn3Sn and Mn3Ge %%%%%%%%%%%%%%%
\begin{figure}[tb]
	\includegraphics[width=0.8\linewidth]{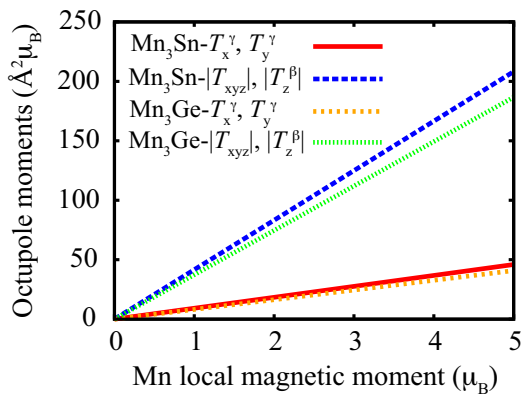}
	\caption{ Local-magnetic-moment dependence of the cluster octupole
 moments for the Mn-clusters in Mn$_3$Sn and Mn$_3$Ge. The
 parameters for the crystal structures are described in Sec.\
 \ref{Sec:Method}.
 $T_{x}^{\gamma}$ and $T_{xyz}$ ($T_{y}^{\gamma}$ and $T_{z}^{\beta}$)
 are for the AFM1 (AFM2) in Fig.\ \ref{Fig:Crystal_Mn3Sn}.
 } 
	\label{Fig:Locmom_CMP}
\end{figure}
%%%%%%%%%%%%%%%%%%%%%%%%%%%%%%%%%%%%%%%%%%%%%%%%%%%%%%%%%
 Next, we identify the CMP moments which characterize the AFM states of
 Mn$_3$$Z$. We here mainly focus on the AFM states of Mn$_3$Sn, whose
 magnetic structures have been well established
 experimentally~\cite{Tomiyoshi1982, Nagamiya1982, Brown1990}.
  Mn magnetic moments in Mn$_3$Sn form the so-called inverse triangular spin
  structure below the N\'{e}el temperature of $T_{\rm
 N1}\simeq$420K~\cite{Tomiyoshi1982, Brown1990}
 (Fig. \ref{Fig:Crystal_Mn3Sn} (b)).
Mn$_3$Sn undergoes another phase transition at $T_{\rm N2}\sim$50K but the
  detailed magnetic structure for the low temperature phase is unknown. 
 We therefore focus on the magnetic phase above 50K.
Interestingly, the inverse triangular spin structure rotates following the
  direction of applied magnetic fields in the
  $c$-plane~\cite{Nagamiya1982}.
  We refer the magnetic structures under the magnetic fields along the $x$
  and $y$ directions as AFM1 and AFM2, respectively, as shown in
  Fig.\ \ref{Fig:Crystal_Mn3Sn} (b).
 The magnetic space groups of the AFM1 and AFM2 magnetic structures
 belong to $Cmc'm'$ and $Cm'cm'$, respectively, taking the primary,
 secondary, and tertiary direction as the $x$, $y$, and $z$ axes.
 The symmetry operators in the magnetic space group are listed in Table \ref{tab:MagSysSym}.
 The magnetic space groups for AFM1 and AFM2 states are the same as those of the
 FM states with the magnetic moments along the $x$ and $y$ directions, respectively.
  Furthermore, these AFM states require the S-O coupling, as well as FM states, to induce
  the AHE from the discussion in Sec. \ref{Sec:SO} on the coplanar spin configuration.

 Because of the geometry of the magnetic alignments on the Mn atoms,
  there is no magnetization of the dipole moment if all the atoms
  have the same size of local magnetic moment~\cite{Note1}. 
 Also, since the magnetic structures preserve the inversion symmetry,
  only the odd rank CMP moments are finite. 
   Indeed, the CMP moments calculated for the
  AFM states with Eq. (\ref{Eq:multipole}) are finite only for the odd
  ranks higher than one. The lowest rank CMP moments characterizing the AFM
  spin configurations are thus cluster octupole moments.

    For the symmetry operators in the $D_{3d}$ point group, the AFM1
  (AFM2) magnetic configuration of Mn$_3$Sn has the same
  transformation property with the magnetic structure characterized by
  $T_x^{\gamma}$ and $T_{xyz}$ ($T_y^{\gamma}$ and $T_z^{\beta}$) in
  Fig.\ \ref{Fig:MagnCMPs}.
  Namely, the magnetic modulation of AFM1 (AFM2) parallel to the $xy$-plane is
  characterized by $T_x^{\gamma}$ ($T_y^{\gamma}$), and the
  three dimensional configuration is characterized by the $T_{xyz}$
  ($T_z^{\beta}$).
  The magnetic configurations in the different clusters are related to
  each other with the operation of $TC_{2z}$
  (See Table \ref{tab:MagSysSym} and Fig.\ \ref{Fig:Crystal_Mn3Sn}).
  The operation of $TC_{2z}$ preserves $T_x^{\gamma}$ ($T_y^{\gamma}$)
  and flips the sign of $T_{xyz}$ ($T_z^{\beta}$) due to the
  transformation property for each IREP. As a result,
   $T_x^{\gamma}$ and $T_{xyz}$ ($T_y^{\gamma}$ and $T_z^{\beta}$)
  octupole moments have ferro and antiferro alignments, respectively,
  between the neighboring clusters.
   Therefore, only the $T_x^{\gamma}$ ($T_y^{\gamma}$)-octupole
  moment can have macroscopic magnetization in the AFM1 (AFM2) state.

  In Fig. \ref{Fig:Locmom_CMP}, we show how the cluster octupole
  moments in Mn$_3$Sn and Mn$_3$Ge change as a function of the local
  magnetic moment~\cite{Note3}. Here we assumed that all the Mn sites always have the
  same size of the local magnetic moment and Mn$_3$Ge has the
  same spin configurations with those of Mn$_3$Sn.
  In this situation, we can show from Eq. (\ref{Eq:multipole}) that the
  octupole moments are proportional to the local magnetic moments.
  Thus the local moment can also quantify the symmetry breaking
  associated with the AHE.
 The local magnetic moment, however, cannot characterize general AFM
 orders when inequivalent magnetic atoms have different
 sizes of local moments. 
 On the other hand, even in such cases, the CMP moments work as the
 order parameter quantifying the symmetry breaking.
%In Fig. \ref{Fig:Locmom_CMP}, the difference of lattice constants of
%Mn$_3$Sn and Mn$_3$Ge leads to the difference of the cluster octupole
%moments for the same local moments.

%%%%%% Energy bands of magnetic states of Fe and Mn3Sn (AFM1) %%%%%%%%%%
\begin{figure}[bt]
        \includegraphics[width=1.0\linewidth]{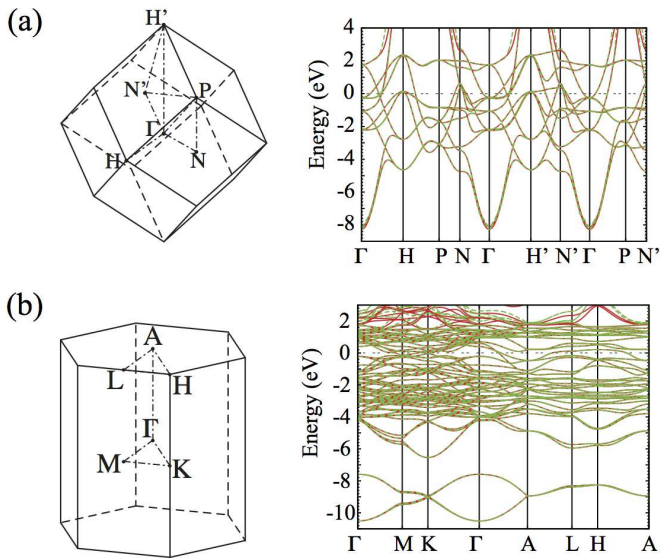}
        \caption{ BRZ and the energy bands for (a) the FM state of
 bcc-Fe and (b) the AFM1 state of Mn$_3$Sn. 
 The red and green lines are the energy bands obtained from the first-principles calculations and from the
 Wannier interpolation, respectively.}
\label{Fig:TBband}
\end{figure}
%%%%%%%%%%%%%%%%%%%%%%%%%%%%%%%%%%%%%%%%%%%%%%%%%%%%%%%%%%%%
%%%%%%%%%%%% AHC of ferromagnetic bcc- Fe%%%%%%%%%%%%%%%
\begin{figure}[bt]
	\includegraphics[width=0.8\linewidth]{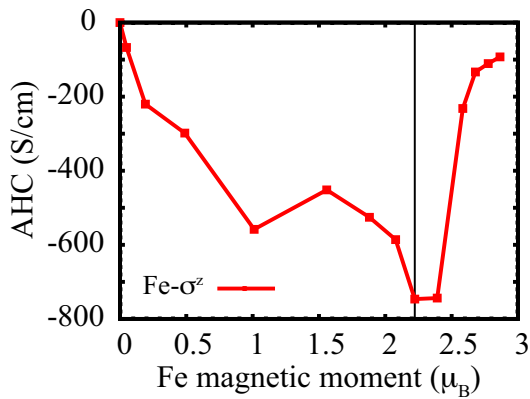}
	\caption{ Local magnetic moment (cluster dipole moment) dependence of the AHC in
 the FM states of bcc-Fe. The solid line in the plot indicates
 the local magnetic moment obtained by the first-principles calculation.
 } 
	\label{Fig:AHC_Fe}
\end{figure}
%%%%%%%%%%%%%%%%%%%%%%%%%%%%%%%%%%%%%%%%%%
%%%%%%%%%%%% DOS and Berry curvature of ferromagnetic bcc-Fe %%%%%%%%%%%% %%%%%%%%%%%%%%%
\begin{figure}[h!]
	\includegraphics[width=0.8\linewidth]{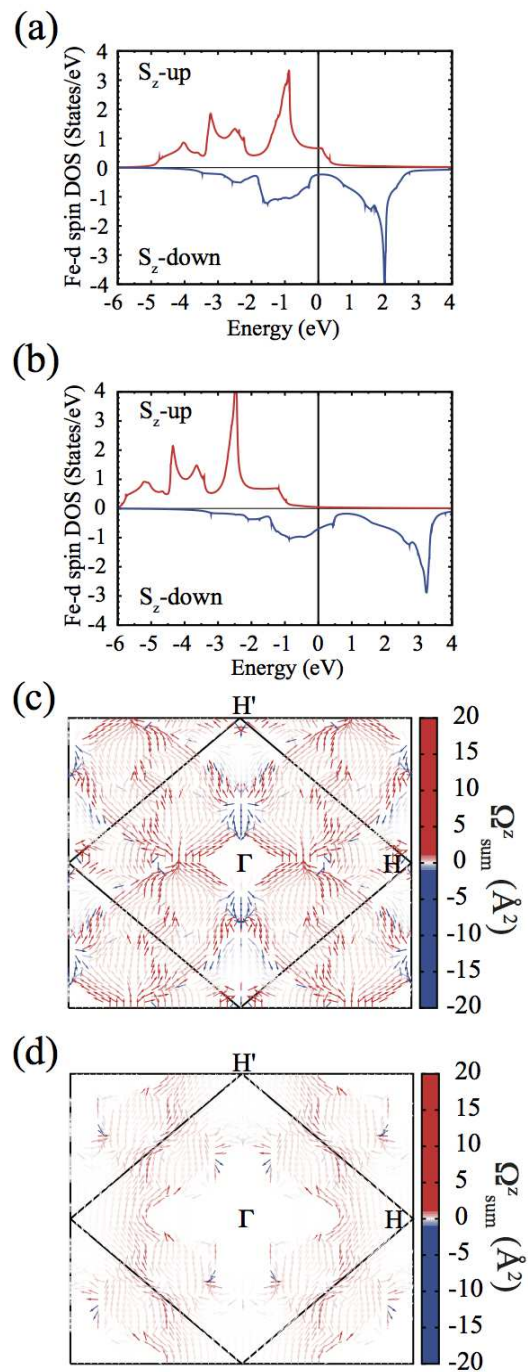}
	\caption{ 
 Fe-$d$ spin-projected DOS for the FM states of bcc-Fe with the magnetic moments
 (a) 2.22 $\mu_B$ and (b) 2.86 $\mu_B$ corresponding to
 $\lambda$= 1.0, and 2.0, respectively. (c) and (d) show 
 ${\mathbf \Omega}_{\rm sum}({\bm k})$ (see text) on the $k_y=0$ plane
 corresponding to the states of (a) and (b), respectively. 
 ${\mathbf \Omega}_{\rm sum}$ vectors are colored by the weight
 of $\Omega^{z}_{\rm sum}$, which contribute to the AHC.
}
	\label{Fig:DOS_Berry_Fe}
\end{figure}
%%%%%%%%%%%%%%%%%%%%%%%%%%%%%%%%%%%%%%%%%%
\section{First-principles analysis of anomalous Hall effect}
\label{Sec:FirstPrinciples}
\subsection{Method}
\label{Sec:Method}
We performed the first-principles calculations for the
AFM1 and AFM2 states of Mn$_3$$Z$ ($Z$=Sn, Ge) with the QUANTUM ESPRESSO
package\cite{Giannozzi2009} with the relativistic version of the
ultrasoft pseudo potentials using the exchange-correlation functional of
the generalized gradient approximation (GGA)
proposed by Perdew, Burke, and Ernzerhof~\cite{Perdew1996}.
 We used the lattice constants $a$=5.665${\rm \AA}$, $c$=4.531${\rm \AA}$ and the Wyckoff position of the
Mn 6$h$ atomic sites $x$=0.8388 from the experimental results\cite{Tomiyoshi1982a,Brown1990}.
For Mn$_3$Ge, the lattice constants $a$=5.34${\rm \AA}$, $c$=4.31${\rm \AA}$ from the experiment were adopted~\cite{Kiyohara2016} and the Wyckoff
parameter of the Mn 6$h$ atomic sites was taken the same as that of
Mn$_3$Sn.
The spin configuration was set such as Fig. \ref{Fig:Crystal_Mn3Sn} (b)
for the AFM1 and AFM2 spin configurations. In the GGA calculations, we
obtained the local magnetic moment 3.39$\mu_{B}$ for Mn$_3$Sn and
2.92$\mu_{B}$ for Mn$_3$Ge both for the AFM1 and AFM2 states.
 The calculations were performed also for bcc Fe as a reference FM system.
 The lattice constant 2.87${\rm \AA}$ was used. The spin moment was set
 to the $+z$ direction. The magnetic moment obtained from the GGA
 calculation is 2.22$\mu_{B}$.

 The realistic tight-binding models were obtained from the
 first-principles band structures, using the Wannier90 program code~\cite{Mostofi2008}.
 The tight-binding model for bcc Fe was generated with the 18 orbitals
 of Fe-$s$, $p$, $d$ orbitals, and Mn$_3${\it Z} with the 88 orbitals using
 Mn-$s$, $d$ and {\it Z}-$s$, $p$ orbitals.
 The energy band structures of the tight-binding models show good
 agreement with those of the first-principles calculations
  as shown in Fig.\ \ref{Fig:TBband}.
  The Berry curvature and AHC were calculated within
  the tight-binding models with Eqs. (\ref{Eq:sigma}), (\ref{Eq:omega}) and (\ref{Eq:verocity}).

  To discuss the magnetic-moment dependence of the AHC, we also
  performed the calculations for nonmagnetic states with the S-O coupling
  and obtained the tight-binding Hamiltonian
  $H_{\rm nmag}$ as well as the Hamiltonian for the magnetic state with
  the S-O coupling, $H_{\rm mag}$.
 We further generated the hopping matrices obtained by interpolating or extrapolating these tight-binding Hamiltonian matrices such as $H_{\lambda} =H_{\rm
 nmag}+\lambda (H_{\rm mag}-H_{\rm nmag})\ (\lambda \geq 0)$. 
 The Fermi level was determined so as to preserve the electron number
 for each $H_{\lambda}$,
  and then the AHC was calculated for
  the obtained electronic structure and the Fermi level.

%%%%%%%%%%%% AHC of antiferromagnetic Mn3Z %%%%%%%%%%%%%%%
\begin{figure}[tb]
	\includegraphics[width=0.8\linewidth]{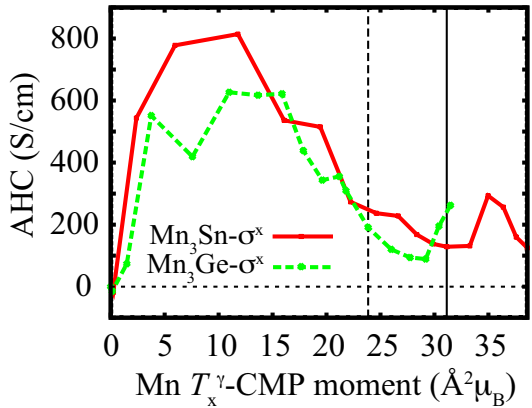}
	\caption{  Mn cluster
 $T_{x}^{\gamma}$-octupole moment dependence of the AHC in AFM1 states of
 Mn$_3$Sn and Mn$_3$Ge. The solid and dashed lines in the plot indicate
 the local magnetic moments obtained by the first-principles calculations for Mn$_3$Sn and Mn$_3$Ge, respectively, obtained by the first-principles calculations.
 } 
	\label{Fig:AHC_Mn3Z}
\end{figure}
%%%%%%%%%%%%%%%%%%%%%%%%%%%%%%%%%%%%%%%%%%%%%%%%%%%%%%%%%%%
%%%%%%%%%%%% DOS and Berry curvature of antiferromagnetic Mn3Z	%%%%%%%%%%%% %%%%%%%%%%%%%%%
 \begin{figure}[h!]
	\includegraphics[width=0.8\linewidth]{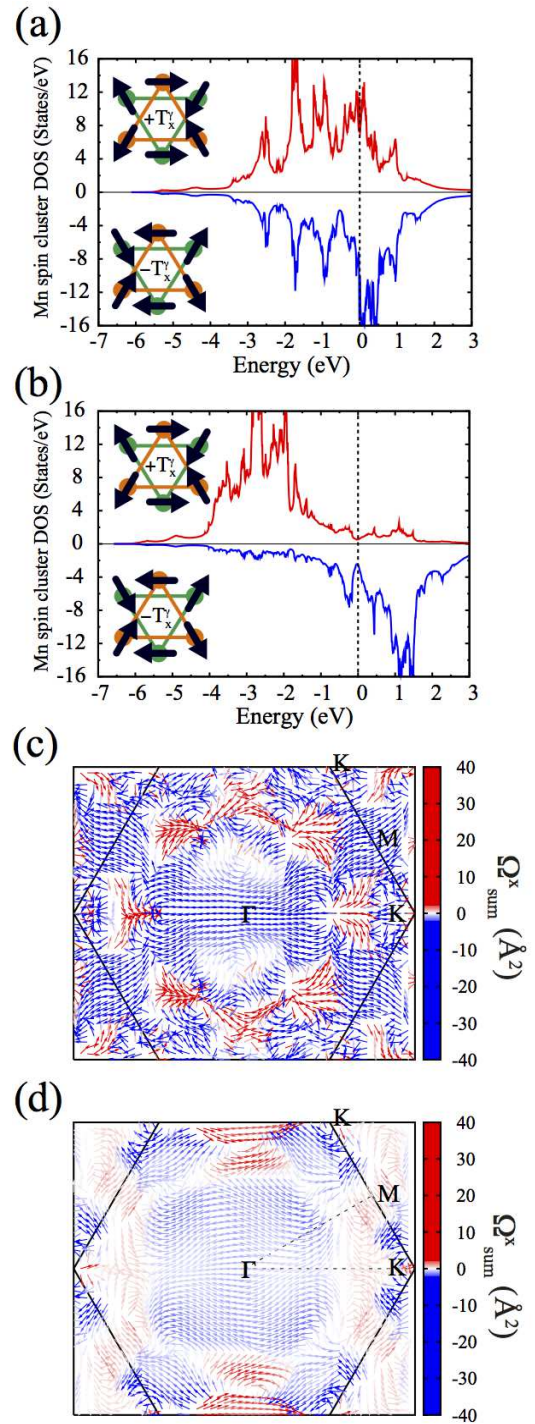}
	\caption{ 
Projected DOS for the AFM states of Mn$_3$Sn with
 the Mn $T_x^{\gamma}$-CMP (local magnetic moment) (a) 11.8${\rm \AA}^2\mu_B$
 (1.28$\mu_B$) and (b) 31.2${\rm \AA}^2\mu_B$ (3.39$\mu_B$) corresponding to
 $\lambda$=0.20, 1.0, respectively.
(c) and (d) show ${\mathbf \Omega}_{\rm sum}({\bm k})$ (see
 text) on the $k_z=0$ plane corresponding to the states of (a) and (b),
 respectively.
 ${\mathbf \Omega}_{\rm sum}$ vectors are colored by the weight of $\Omega^{x}_{\rm sum}$, which contribute to the AHC.
 } 
	\label{Fig:DOS_Berry_Mn3Z}
 \end{figure}
%%%%%%%%%%%%%%%%%%%%%%%%%%%%%%%%%%%%%%%%%%%%%%%%%%%%%%%%%%%
\subsection{Ferromagnetic states of bcc Fe}
\label{Sec:Ferro}
 Before proceeding to the first-principles analysis of the AFM states of
 Mn$_3$$Z$, we discuss the AHE in the FM states of bcc Fe, which has
 well been investigated theoretically~\cite{Fang2003,
 Yao2004, Wang2006, Gosalbez2015}.
  Figure \ref{Fig:AHC_Fe} shows the magnetic moment (cluster dipole moment) dependence of the AHC for bcc Fe.
  The AHC increases as the magnetic moment increases in the small
 magnetic moment region and  makes a peak around 2.22$\mu_B$ obtained in the
 GGA calculation, and then decreases.
  Figures \ref{Fig:DOS_Berry_Fe} (a) and (b) show spin density of states
 (DOS) of the Fe-$d$ orbitals for the electronic structures with the
 magnetic moments 2.22$\mu_B$, and 2.86$\mu_B$, respectively.
  The original magnetic state with the magnetic moment 2.22$\mu_B$ has
 the contribution from both the majority- (up-) and minority- (down-)spin DOS at the
 Fermi level. Meanwhile, the DOS of Fig. \ref{Fig:DOS_Berry_Fe}
 (b) shows that the up-spin states are almost fully occupied due to the large
 spin moment and do not have weight at the Fermi level.
  Figures \ref{Fig:DOS_Berry_Fe} (c) and (d) show the ${\bm k}$-dependence of the
 Berry curvature summed over the occupied states, ${\mathbf
 \Omega}_{\rm sum}({\bm k})=\sum_{n}f(\varepsilon_n({\bm
k})-\mu){\mathbf \Omega}_{n}({\bm k})$, on the $k_y=0$ plane.
 In both states, $\Omega^z_{\rm sum}$ is positive in a large region of the BRZ, leading to
 negative $\sigma^z$ by the BRZ integration. Meanwhile, we see
 that the intensity of $\Omega^{z}_{\rm sum}({\bm k})$ is much stronger
 in Fig. \ref{Fig:DOS_Berry_Fe} (c) than in Fig. \ref{Fig:DOS_Berry_Fe} (d).

\subsection{Antiferromagnetic states of Mn$_3$Sn}
%%%
% Band structure of Mn3Sn
%%%
 Let us move on to Mn$_3$$Z$. We here mainly focus on the AFM1 state of Mn$_3$Sn, whose
 crystal and magnetic structures have been well established
 experimentally~\cite{Tomiyoshi1982, Nagamiya1982, Brown1990}.
  The local Mn magnetic moment 3.39$\mu_B$ and $\sigma_{yz}$=129(S/cm) obtained from the GGA calculation agree well with
 the experimental measurement of the local magnetic moment ~3$\mu_B$ and that of
 the AHC $\sim$100(S/cm)~\cite{Nakatsuji2015}.
 The calculated value is also consistent with a recent work~\cite{Zhang2016}.
   Figure \ref{Fig:AHC_Mn3Z} shows that the AHC in Mn$_3$Sn and
 Mn$_3$Ge shows a similar CMP-moment dependence. 
  The calculations also show that decreasing the magnetic moments from
  the one obtained with GGA makes the size of the AHC larger. 
  In fact, for the electronic structure obtained by the GGA calculation,
 the AHC (magnetization) of Mn$_3$Ge is larger (smaller) than that of Mn$_3$Sn.
  This is consistent with the recent
 experiments~\cite{Nakatsuji2015, Nayak2016, Kiyohara2016}.

% For collinear FM states, the spin is a good quantum number when the S-O
%   interaction is neglected, and the S-O interaction couples the two
%   different spin states.
   For the AFM states of Mn$_3$Sn, we define the Mn cluster spin bases
 as the two symmetrized spin configurations related by the time reversal symmetry
 (see the insets of Fig. \ref{Fig:DOS_Berry_Mn3Z} (a) and (b)). The
 two spin configurations are characterized by the positive and negative
 $T_{x}^{\gamma}$ octupole moments.
 Then, we can discuss the AHE in terms of the spin cluster,
 in analogy with the majority- and minority-spin states in the FM systems.
 Figures \ref{Fig:DOS_Berry_Mn3Z} (a) and (b) show the projected DOS for 
 each spin cluster corresponding to the
 $T_x^{\gamma}$= 11.8${\rm \AA}^2\mu_B$ and 31.2${\rm \AA}^2\mu_B$, respectively.
  The magnetic state with $T_x^{\gamma}$=11.8${\rm \AA}^2\mu_B$ has a large DOS contribution from the both spin cluster components at the Fermi level. 
 On the other hand, in the magnetic state with the large $T_x^{\gamma}$
 moment, the spin cluster states characterized by the positive
 $T_x^{\gamma}$ are almost fully occupied due to the large octupole
 moment and only have small weight at the Fermi level.
   Figures \ref{Fig:DOS_Berry_Mn3Z} (c) and (d) show ${\mathbf
 \Omega}_{\rm sum}({\bm k})$ colored by the $\Omega^x_{\rm sum}$ component on the $k_z=0$ plane. In both states, $\Omega^x_{\rm sum}$ is negative in a large region of the BRZ, leading to
 positive $\sigma^x$ by the BRZ integration.
  The intensity of $\Omega^x_{\rm sum}({\bm k})$ is stronger in Fig.\
 \ref{Fig:DOS_Berry_Mn3Z} (c) than in Fig.\ \ref{Fig:DOS_Berry_Mn3Z}
 (d), which is a situation similar to that in
 Fig. \ref{Fig:DOS_Berry_Fe} (c) and (d).
%  The Fermi surfaces are formed by the 50th, 51st, and 52nd energy bands
% in our TB models for the AFM states of Mn$_3$Sn.

\section{Summary}
\label{Sec:Summary}
   We showed that the symmetry breaking due to the commensurate non-collinear
 magnetic order can be measured with the CMP moment, which is defined
 for atomic clusters in the crystal.
 We identified the degree of freedom responsible for the AHE in generic
 magnetic systems as the macroscopic contribution of $T_{1g}$ ($A_{2g}$
 or/and $E_{1g}$) CMP moments for the $O_{h}$ ($D_{6h}$) point-group
 representation.
  The theoretical framework was applied to the AFM states of
   Mn$_3$Ir and Mn$_3$$Z$.
   The AFM1 (AFM2) state of Mn$_3$Sn is characterized with the
   $T^{\gamma}_{x}$ and $T_{xyz}$ ($T^{\gamma}_{y}$ and $T_{z}^{\beta}$) cluster octupole moments in the
 $D_{6h}$ IREPs and the AHE is induced by the macroscopic contribution
   of $T^{\gamma}_{x}$ ($T^{\gamma}_{y}$) with the S-O coupling.
The AHC in the FM states of bcc-Fe and that in the
 AFM states of Mn$_3$Sn show similar dependence on the CMP moments.
   Thus, the CMP makes it possible to discuss the FM and AFM
 states in the same framework, and is useful to search for another new functional
 material having a large AHE.

\section*{Acknowledgments}
We thank S. Nakatsuji, H. Kusunose, and T. Miyake for helpful comments and
discussions. 
This work was supported by JSPS KAKENHI Grant Numbers JP15K17713, JP15H05883 (J-Physics),
 JP16H04021, JP16H00924, JP16H06345 and PRESTO and CREST, Japan Science and Technology Agency.

\bibliographystyle{revtex}
\bibliography{AHE_AFM}

\end{document}